\documentclass[preprint,12pt,authoryear]{elsarticle}

\usepackage{amssymb}
\usepackage{amsmath}
\usepackage{listings}
\usepackage{xcolor} 
\journal{Computer Networks}

\begin{document}

\begin{frontmatter}



\title{Mazzaroth: A High-Throughput DAG Consensus with State Root} 


\author{Haohan Li} 

\affiliation{organization={mazzarothnet},
            addressline={ferteammarch@gmail.com}}

\begin{abstract}
 \cite{bitcoin2008} Nakamoto Consensus achieves a decentralized ledger through a single-chain blockchain, assuming a maximum network delay, which limits block generation speed, resulting in low throughput. \cite{pg2018} (PG) enhances throughput using a blockDAG structure, but its probabilistic confirmation restricts smart contract applications. To address this, Mazzaroth proposes a Pow-based blockDAG consensus, employing a linear ordering algorithm to compute the \cite{eth} and achieve state finality, thereby supporting smart contracts. Its dynamic difficulty adjustment, independent of the assumption, adapts to network and hashrate fluctuations, ensuring state consistency via a head chain while maximizing throughput. Simulations validate Mazzaroth's efficient consensus performance. This paper presents the Mazzaroth ordering algorithm, the difficulty adjustment mechanism, and performance evaluation.
\end{abstract}


\end{frontmatter}



\section{INTRODUCTION}
\label{sec1}

The Bitcoin Nakamoto consensus achieves a decentralized ledger through a single-chain blockchain, assuming an upper bound on the network diameter delay $ \Delta $ (the maximum propagation delay between any two nodes) and setting an expected block generation time $ \lambda $ significantly greater than $ \Delta $ to ensure block propagation. However, this restricts block generation speed, resulting in low transaction throughput and poor adaptability to network delay fluctuations. Reducing $ \lambda $ compromises decentralization, while increasing $ \lambda $ decreases throughput, creating a trade-off dilemma.
PHANTOM GHOSTDAG (PG) employs a blockDAG structure to enhance throughput, with its algorithm addressing consensus through transaction ordering. However, its probabilistic confirmation, complex state computation, and lack of dynamic difficulty adjustment limit smart contract applications.
Mazzaroth proposes a Pow-based blockDAG consensus, achieving rapid state finality through a sorting algorithm to support smart contracts. Its dynamic difficulty adjustment, independent of the assumption $ \Delta $, adapts to network fluctuations, utilizing a head chain to form the root of the state and pushing throughput to the limit of the network. This paper first introduces Mazzaroth's sorting algorithm, followed by performance test metrics that guide the presentation of its difficulty adjustment mechanism.

\section{Mazzaroth Protocol: Sorting Algorithm}

\subsection{Preliminaries}
\label{subsec1}

This paper employs the following terminology. A directed acyclic graph (DAG) is denoted as $ G = (C, E) $, where $ C $ represents blocks, and $ E $ represents hash references between blocks. We frequently write $ B \in G $ instead of $ B \in C $.

$ \text{parent}(B, G) \subset C $ denotes the set of parent nodes of block $ B $.

$ \text{past}(B, G) \subset C $ denotes the subset of blocks reachable from $ B $ (excluding $ B $ itself).

$ \text{size}(B, G) \subset C $ denotes the number of blocks in $ \text{past}(B, G) $ plus 1 (including $ B $ itself).

$ \text{anticone}(B, G) \subset C $ denotes the set of nodes not in $ \text{past}(B, G) $.

$ \text{tips}(G) \subset C $ denotes the leaf nodes, i.e., blocks with an in-degree of 0.

It should be noted that once the block $ B $ is mined, $ \text{past}(B, G) $ and $ \text{size}(B, G) $ are fixed, and no subsequent actions by honest or malicious miners can alter them.
\begin{figure}[t]
\centering
\includegraphics{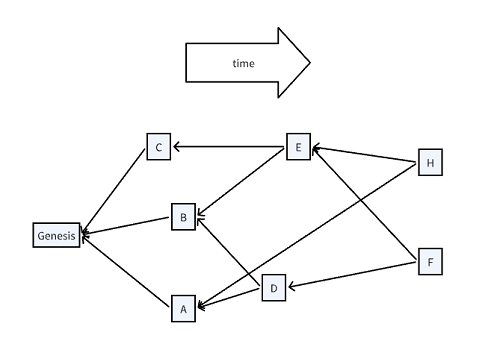}
\caption{$ \text{parent}(F, G) = {E, D} $
$ \text{past}(F, G) = {\text{Genesis}, A, B, C, D, E, F} $
$ \text{size}(F, G) = 7 $
$ \text{anticone}(F, G) = {H} $
$ \text{tips}(G) = {H, F} $}\label{fig1}
\end{figure}

\subsection{Mazzaroth Sorting Protocol}
Similar to the PHANTOM protocol, Mazzaroth incorporates both blocks containing conflicting transactions into the blockDAG, ensuring that all miners reference them.
Broadly, the Mazzaroth sorting process comprises the following steps:
Identify the block with the maximum size(B) among tips(G) and designate it as the Head. Recursively compute the ordering for past(Head). For blocks not included in past(Head), respect topological ordering and sort based on block size, then append these blocks sequentially after the past(Head) ordering. Output the finalized transaction ordering.
\subsubsection{Formalized Sorting Algorithm}
The procedure outlined above is formalized in the algorithm in the following. The algorithm takes a BlockDAG as input and outputs an ordered list containing all blocks in G. Initializes with a DAG that contains only the genesis block (lines 2-3). It then examines the past of each leaf block (tips) to identify the block with the maximum size(B, G), inheriting its ordering (lines 4-7). Subsequently, it applies topological sorting to place poorly connected nodes at the end (lines 8-11), adds the current node to the ordering, and finally merges the results (line 12).

\begin{lstlisting}[language=Python, caption={Ordering the DAG}, label={lst:sorting}]
function OrderDAG(N,G)
	if G == {genesis} then
		return [genesis]
	for B in parent(N,G) do
		[SetSize] <- size(B,G)
	Bmax <- argmax(SetSizeB: B in tips(G))
	OrderedList <- OrderDAG(Bmax,past(Bmax))
	EndList <- []
	for B in anticone(Bmax,G) Using topological order, prioritizing nodes with a smaller $ \text{size}(B, G) $.
		add B to the front of EndList
		add EndList to the end of OrderedList
	add N to OrderedList
	return OrderedList
\end{lstlisting}

\section{Metrics}
\subsection{Transactions}
We define a transaction as primarily composed of the following components (omitting information not related to the sorting algorithm).

Inputs: Balances of multiple accounts.
Outputs: Balances of multiple accounts (including all input accounts, even if the balance is zero at that time).
Signatures: Signatures of each account with a reduced balance.
Constraints: The total balance of inputs equals the total balance of outputs.
The validity of a transaction depends on whether the input account balances align with the transaction's claims. Accounts may set a minimum received balance to mitigate DDoS attacks. Based on the transaction sequence generated by the sorting algorithm, Mazzaroth sequentially traverses transactions to compute the global state root. The Mazzaroth virtual machine supports state rollback: when processing forward, a transaction is accepted if its input balance is valid, otherwise it is ignored; when processing backward, a transaction is accepted if its output balance is valid, otherwise it is ignored. This mechanism enables the consensus algorithm to derive the state root of previous blocks from the current block's state root, making it possible to transition state roots between different blocks in the blockDAG.

\subsection{Parameters and Testing}
The following metrics are used to evaluate the consensus performance of Mazzaroth:

$ \text{head}(B, G) $: The block with the maximum $ \text{size} $ among $ \text{parent}(B, G) $.

$ \text{link}(B, G) $: The set of blocks was traced upward from $ \text{head}(B, G) $, that is, $ \text{head}(B, G) $, $ \text{head}(\text{head}(B, G), G) $, $ \text{head}(\text{head}(\text{head}(B, G), G), G) $,..., $ \text{Genesis} $.

$ \text{agree}({B, P}, G) $: The block with the maximum $ \text{size} $ in the intersection of $ \text{link}(B, G) $ and $ \text{link}(P, G) $.

$ \text{distance}(B, G) $: The difference between the $ \text{size} $ of $ B $ and the $ \text{size} $ of the $ \text{agree} $ of its parents, i.e., $ \text{size}(B, G) - \text{size}(\text{agree}(\text{parent}(B, G), G), G) $.

$ \text{state}(B, G) $: The set formed by $ \text{past}(B, G) $ and block $ B $, where the set is ordered and used to compute the state root of the Mazzaroth virtual machine.

In the sorting algorithm, $ \text{state}(B, G) $ is inherited from $ \text{state}(\text{head}(B, G), G) $. Furthermore, the $ \text{state} $ of all nodes in $ \text{parent}(B, G) $ can be mutually derived. When $ \text{distance}(B, G) $ is small, the network reaches consensus; when $ \text{distance}(B, G) $ is large (e.g., $ \text{agree}(\text{tips}(G), G) = \text{genesis} $), it indicates consensus failure, potentially leading to network partitioning. The number of blocks mined within the dynamic network diameter delay $ D' $ (distinguished from the fixed upper bound $ \Delta $) is defined as Block Per Delay (BPD), and $ \text{distance}(B, G) $ is abbreviated as DIS. BPD and DIS are positively correlated.

The simulation test utilized 1000 miners (900 synchronized within $ D' $, 100 with delayed synchronization), mining a total of 100,000 blocks, and recorded the variation of DIS with different BPD values (code available at \cite{mazzaroth}). Figure 2 illustrates the trend of DIS increasing as BPD rises. Figure 3 shows the difference between the $ \text{size} $ of $ \text{agree}(\text{tips}(G), G) $ and the maximum $ \text{size} $ in $ \text{tips}(G) $, revealing that when $BPD > 14$, $ \text{agree} $ is often $ \text{genesis} $, indicating a consensus split.

\begin{figure}[t]
\centering
\includegraphics{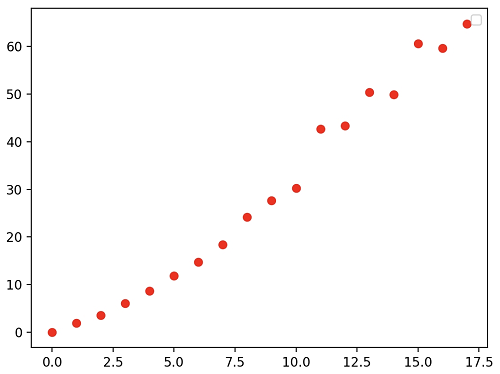}
\caption{ This figure illustrates the variation of DIS across different BPD values, reflecting the stability of network consensus. }\label{fig2}
\end{figure}

\begin{figure}[t]
\centering
\includegraphics{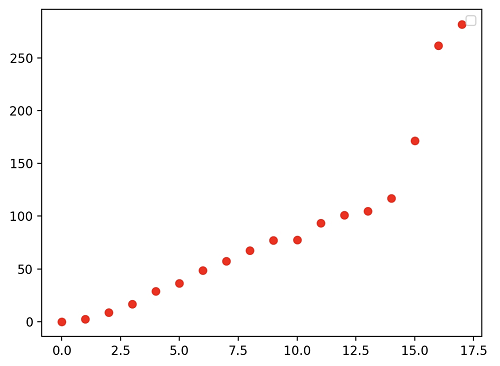}
\caption{ This figure depicts the difference between the $ \text{size} $ of $ \text{agree}(\text{tips}(G), G) $ and the maximum $ \text{size} $ in $ \text{tips}(G) $ across varying BPD values. }\label{fig3}
\end{figure}

\section{Mazzaroth Protocol: Difficulty Adjustment Algorithm}
The difficulty adjustment algorithm is equally critical as the sorting algorithm, together forming the core mechanisms of Mazzaroth. This section elucidates how Mazzaroth optimizes the blockDAG structure through dynamic difficulty adjustment, facilitating efficient state root computation and maximizing throughput.

\subsection{Design Philosophy}

Bitcoin's Nakamoto Consensus regulates block generation time (approximately 10 minutes) through difficulty adjustment, assuming a maximum network delay $ \Delta $, ensuring block propagation and the formation of a single-chain structure, where the state root on the chain represents consensus. However, this mechanism constrains throughput. Mazzaroth, built upon a blockDAG structure, aims to establish a stable sequence for the head chain, where the state root of the longest head chain signifies consensus, equivalent to maintaining $ \text{distance}(B, G) $ (DIS) within a bounded range. Testing results (see Section 3.2) demonstrate that DIS is positively correlated with Block Per Delay (BPD, i.e., the number of blocks mined within the dynamic network delay $ D' $). Mazzaroth dynamically adjusts the Proof of Work (PoW) difficulty by observing DIS, eliminating the need to record timestamps, thereby optimizing the blockDAG structure.

\subsection{Algorithm Description}

Mazzaroth's difficulty adjustment is based on a fitted function $ f(\text{DIS}) = \text{BPD} $, which captures the correspondence between DIS and BPD. The algorithm establishes an expected BPD (PBPD), representing the target block generation rate under varying network delays and computational power. The PoW difficulty is defined by $ \text{SHA256} < \text{target} $, with the adjustment formula as follows:

\begin{equation}
target=target' * f(DIS') / PBPD
\end{equation}

where $ \text{target}' $ is the previous difficulty, $ \text{DIS}' $ is the currently observed DIS, and $ \text{PBPD} $ is the expected BPD.

Mazzaroth employs two difficulty targets:

target1: Governs long-term fluctuations, primarily driven by the network's total computational power, adjusted every $ 2016 \times 10 \times 60 $ blocks, based on the average block difficulty and DIS.
target2: Manages short-term fluctuations, primarily influenced by network delay, adjusted for each block based on the DIS and average difficulty of its parent.
A block's SHA256 hash must satisfy both $ \text{target1} $ and $ \text{target2} $ to be accepted.

\subsection{Application Scenarios}
PBPD reflects the trade-off between consensus speed and throughput. A smaller PBPD reduces DIS, accelerating transaction confirmation, which is suitable for low-latency scenarios; a larger PBPD increases DIS, enhancing throughput, which is ideal for high-load applications. In practical deployments, PBPD can be tuned to optimize Mazzaroth's performance according to specific requirements.

\section{Conclusion}
Mazzaroth introduces a blockDAG-based consensus protocol that addresses the challenge of efficiently computing the state root in traditional blockDAGs by designing a novel sorting algorithm and a dynamic difficulty adjustment mechanism. The sorting algorithm generates a transaction sequence through linear ordering, enabling rapid state root computation; the difficulty adjustment algorithm optimizes the blockDAG structure by controlling Block Per Delay (BPD), maintaining a bounded $ \text{distance} $ (DIS).

Simulation tests validate Mazzaroth's high efficiency, demonstrating its ability to maximize throughput at high BPD values while ensuring rapid transaction confirmation and state finality (see Section 3.2). Under the assumption that honest nodes control the majority of computational power, Mazzaroth achieves high-throughput consensus, supporting smart contract applications.

\end{document}